\documentclass[twocolappendix]{emulateapj}

\usepackage{apjfonts,color,amsmath,textcomp,url,graphicx,subfigure}

\newcommand{\kms}{\hbox{km\,s$^{-1}$}}
\newcommand{\cmss}{\hbox{cm\,s$^{-2}$}}
\newcommand{\MJup}{$M_{\mathrm{Jup}}$}
\newcommand{\LBol}{$L_{\mathrm{Bol}}$}
\newcommand{\RJup}{$R_{\mathrm{Jup}}$}
\newcommand{\masyr}{$\mathrm{mas}\,\mathrm{yr}^{-1}$}
\newcommand{\Teff}{\ensuremath{T_{\mathrm{eff}}}}

\slugcomment{To appear in ApJ letters.}

\shorttitle{A New T Dwarf Member of AB~Doradus}
\shortauthors{Gagn\'e et al.}

\begin{document}

\title{SDSS~J111010.01+011613.1: A NEW PLANETARY-MASS T DWARF MEMBER OF THE AB~DORADUS MOVING GROUP}

\author{Jonathan Gagn\'e\altaffilmark{1},\, Adam J. Burgasser\altaffilmark{2},\, Jacqueline K. Faherty\altaffilmark{3,4,5},\, David Lafreni\`ere\altaffilmark{1},\, Ren\'e Doyon\altaffilmark{1},\, Joseph C. Filippazzo\altaffilmark{4,6,7},\, Emily Bowsher\altaffilmark{8},\, Christine P. Nicholls\altaffilmark{9,10}}
\affil{\altaffilmark{1} Institut de Recherche sur les Exoplan\`etes (iREx), Universit\'e de Montr\'eal, D\'epartement de Physique, C.P.~6128 Succ. Centre-ville, Montr\'eal, QC H3C~3J7, Canada jonathan.gagne@astro.umontreal.ca}
\affil{\altaffilmark{2} Center for Astrophysics and Space Sciences, University of California, San Diego, 9500 Gilman Dr., Mail Code 0424, La Jolla, CA~92093, USA}
\affil{\altaffilmark{3} Department of Terrestrial Magnetism, Carnegie Institution of Washington, Washington, DC~20015, USA}
\affil{\altaffilmark{4} Department of Astrophysics, American Museum of Natural History, Central Park West at 79th Street, New York, NY~10024, USA}
\affil{\altaffilmark{5} Hubble Fellow}
\affil{\altaffilmark{6} Department of Engineering Science and Physics, College of Staten Island, City University of New York, 2800 Victory Blvd, Staten Island, NY~10314, USA}
\affil{\altaffilmark{7} The Graduate Center, City University of New York, New York, NY~10016, USA}
\affil{\altaffilmark{8} Department of Astronomy, Columbia University, 550 West 120th Street, New York, NY 10027, USA}
\affil{\altaffilmark{9} Institute for Astrophysics, University of Vienna, Tuerkenschanzstrasse 17, 1180 Vienna, Austria}
\affil{\altaffilmark{10} Lise Meitner Fellow}

\begin{abstract}

We present a new radial velocity measurement that, together with a trigonometric parallax, proper motion and signs of low gravity from the literature, confirms that SDSS~J111010.01+011613.1 is a new T5.5 bona fide member of AB~Doradus. Fitting $\lambda/\Delta\lambda$ $\approx$ 6000 FIRE spectroscopy in the 1.20--1.33\,$\mu$m region to BT-Settl atmosphere models yielded a radial velocity of $7.5 \pm 3.8$\,\kms. At such a young age (110--130\,Myr), current evolution models predict a mass of $\sim$\,10--12\,\MJup, thus placing SDSS~J1110+0116 well into the planetary-mass regime. We compare the fundamental properties of SDSS~J1110+0116 with a sequence of seven recently identified M8--T5 brown dwarf bona fide or high-confidence candidate members of AB~Doradus. We also note that its near-infrared $J-K$ color is redder than field T5--T6 brown dwarfs, however its absolute $J$-band magnitude is similar to them. SDSS~J1110+0116 is one of the few age-calibrated T dwarfs known to date, as well as one of the coolest bona fide members of a young moving group.
\end{abstract}

\keywords{brown dwarfs --- stars: kinematics and dynamics --- techniques: radial velocities}

\section{INTRODUCTION}

Young, low-mass brown dwarfs of the solar neighborhood provide a unique opportunity to understand the properties of directly-imaged gaseous giant exoplanets and study the low-mass end of the initial mass function. Young moving groups in the solar neighborhood are ideal laboratories to search for such young brown dwarfs, due to their relative proximity to the Sun ($\lesssim$\,100\,pc) and their young age ($\lesssim$\,120\,Myr). These moving groups consist of stars that have formed together at the same time and in a common environment, and that share similar space velocities (e.g., see \citealp{2004ARA&A..42..685Z}).

The first few brown dwarf members and candidate members of these young moving groups were identified both in pointed searches and as serendipitous discoveries (e.g., \citealp{1998Sci...282.1309R,2002ApJ...575..484G,2010ApJ...715L.165R,2013AJ....145....2F,2013ApJ...777L..20L,2014AJ....147...34S,2014ApJ...787....5N,2015ApJ...799..203G}). The main difficulties that hindered locating such low-mass members of moving groups were intrinsic faintness and sparse location on the sky. Trigonometric distances and radial velocities are not available for all known brown dwarfs that could correspond to low-mass members of young moving groups. Furthermore, the census of brown dwarfs in the solar neighborhood is still incomplete (e.g., \citealp{2013ApJ...767L...1L,2014A&A...561A.113S}).

T dwarf members younger than $\sim$\,120\,Myr have masses in the planetary regime and are located in the immediate solar neighborhood ($\lesssim$\,20\,pc), which makes them accessible now for follow-up observing and prime targets for detailed study using the next generation of facilities such as the James Webb Space Telescope \citep{2006SSRv..123..485G},  the Thirty Meter Telescope \citep{2008SPIE.7012E..1AN}, the European Extremely Large Telescope \citep{2008SPIE.7012E..19G} and the Giant Magellan Telescope \citep{2014SPIE.9145E..1FJ}. To date, only one isolated T dwarf planetary-mass candidate member of a young moving group has been uncovered, CFBDSIR~J214947.2--040308.9, a T7 candidate member of the $\sim$\,120\,Myr AB~Doradus moving group \citep{2004ApJ...613L..65Z} with an estimated mass of 4--7\,\MJup\ \citep{2012A&A...548A..26D}. However, a parallax measurement has recently rejected a possible membership of CFBDSIR~J214947.2--040308.9 to AB~Doradus or any other known moving group (P. Delorme et al., in preparation). In this Letter we report the identification of a T dwarf bona fide member of AB~Doradus, SDSS~J111010.01+011613.1 (hereafter SDSS J1110+0116).

\section{THE IDENTIFICATION OF SDSS~J111010.01+011613.1}\label{sec:identif}

We have identified SDSS~J1110+0116 as a candidate member of AB~Doradus while preparing the \emph{BASS-Ultracool} survey, which is an extension of the \emph{BASS} survey (\citealt{2015ApJ...798...73G}; Paper~V hereafter) that will aim at the identification of ultracool $\geq$\,L5 brown dwarf members of young moving groups. The survey is based on an updated version of the BANYAN~II tool (\citealt{2014ApJ...783..121G}; Paper~II hereafter). More detail on the \emph{BASS-Ultracool} survey will be presented in a future paper (J.~Gagn\'e et al., in preparation). The BANYAN~II tool uses the sky position, proper motion, the \emph{2MASS} and \emph{AllWISE} photometry, radial velocity and trigonometric distance to derive a membership probability that an object belongs to several young moving groups and the field based on spatial and kinematic models. A detailed description of this tool can be found in Paper~II.

SDSS~J1110+0116 was identified by \cite{2004AJ....127.3553K} as a peculiar T5.5 dwarf that displays unusually strong \ion{K}{1} absorption doublets at $1.17$ and $1.25$\,$\mu$m; these are a sign of low surface gravity as predicted by atmosphere models for T dwarfs (see the discussion in \citealp{2004AJ....127.3553K}). They also noted that its $H-K$ color is unusually red for its spectral type, which further strengthens the low-gravity hypothesis. Using atmosphere models of \cite{2002ApJ...568..335M}, they inferred a surface gravity of $\log g \approx$\,4.0--4.5 and an estimated age of 100--300\,Myr for this object. More recently, \cite{2012ApJS..201...19D} measured a trigonometric distance of $19.19 \pm 0.44$\,pc and a proper motion of $-217.1 \pm 0.7$\,\masyr\ ($\mu_\alpha\cos\delta$) and $-280.9 \pm 0.6$\,\masyr\ ($\mu_\delta$). From our cross-match of the \emph{2MASS} and \emph{AllWISE} catalogs we calculate $\mu_\alpha\cos\delta = -245.0 \pm 19.1$\,\masyr\ and $\mu_\delta = -279.1 \pm 18.8$\,\masyr, which is marginally consistent with (1.6$\sigma$) but less precise than the measurement of \cite{2012ApJS..201...19D}. We thus adopt the former measurement for the remainder of this work. Without information on the radial velocity of this object, the observables mentioned above along with the BANYAN~II tool allowed us to obtain a membership probability of 94.4\% to AB~Doradus and a negligible membership probability to any other moving groups considered.

The predicted radial velocity if this object is a member of AB~Doradus is $4.3 \pm 1.8$\,\kms. We did not include photometric measurements in this calculation since a trigonometric distance is available and the young photometric sequences are still uncertain for such a late spectral type. However, even if we use the extended color-magnitude sequences described in Paper~II, we obtain similar results. The Monte Carlo analysis described in Paper~II yields a small probability (0.6\%) that a relatively young ($<$\,1\,Gyr) object from the field obtains such a high bayesian membership probability to AB~Doradus at this galactic latitude ($b=54.53$\textdegree) and for this magnitude of proper motion ($355$\,\masyr).

The age of AB~Doradus (110--130\,Myr; \citealp{2013ApJ...766....6B}) is consistent with the estimated age of \cite{2004AJ....127.3553K} for this object, further indicating that SDSS~J1110+0116 is a compelling candidate member of AB~Doradus.

\section{OBSERVATIONS}\label{sec:obs}

\begin{figure}
	\centering
	\subfigure[Observed spectrum and best-fitting BT-Settl model]{\includegraphics[width=0.445\textwidth]{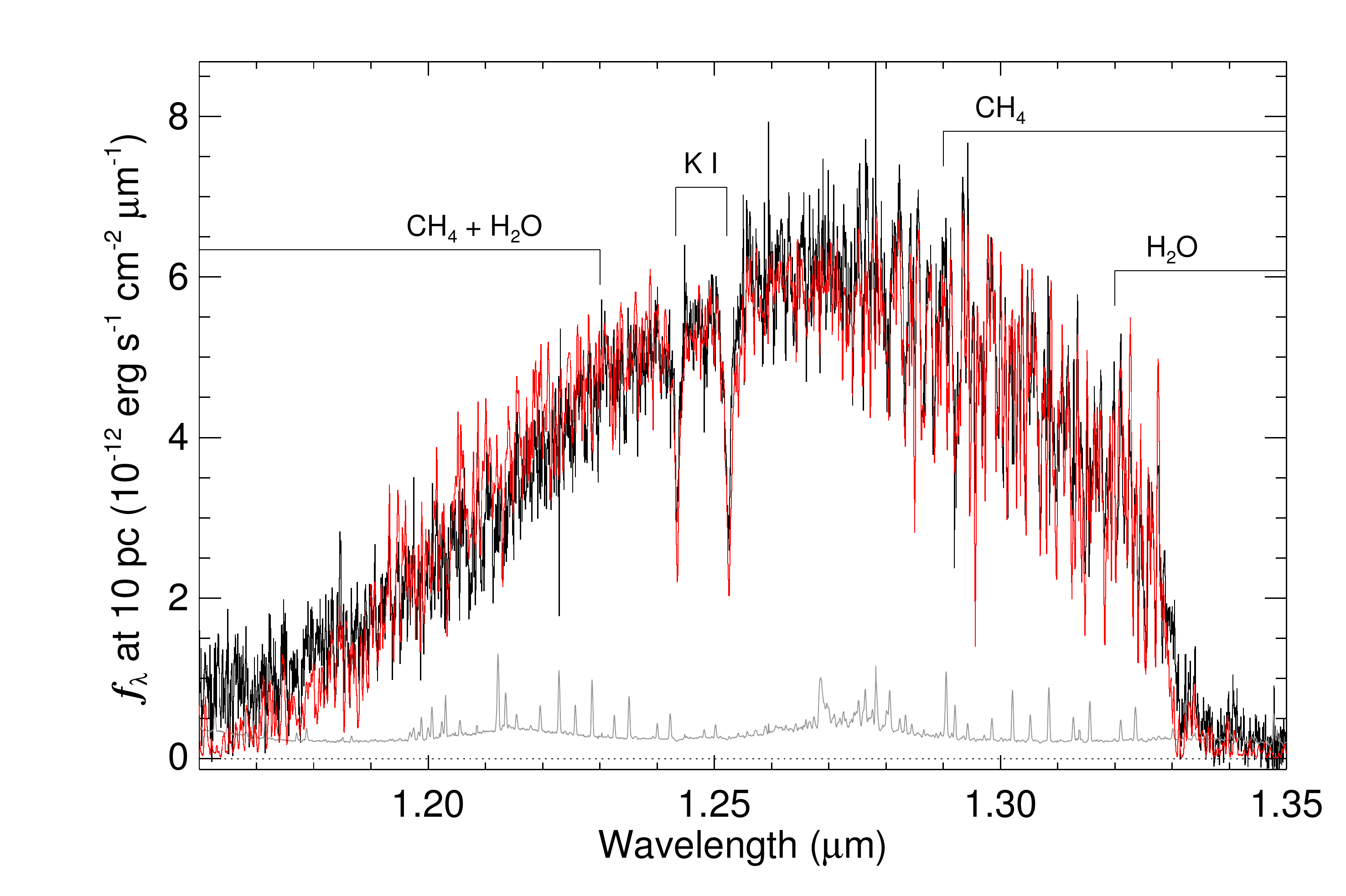}\label{fig:xcorla}}
	\subfigure[Distribution of radial velocity measurements]{\includegraphics[width=0.445\textwidth]{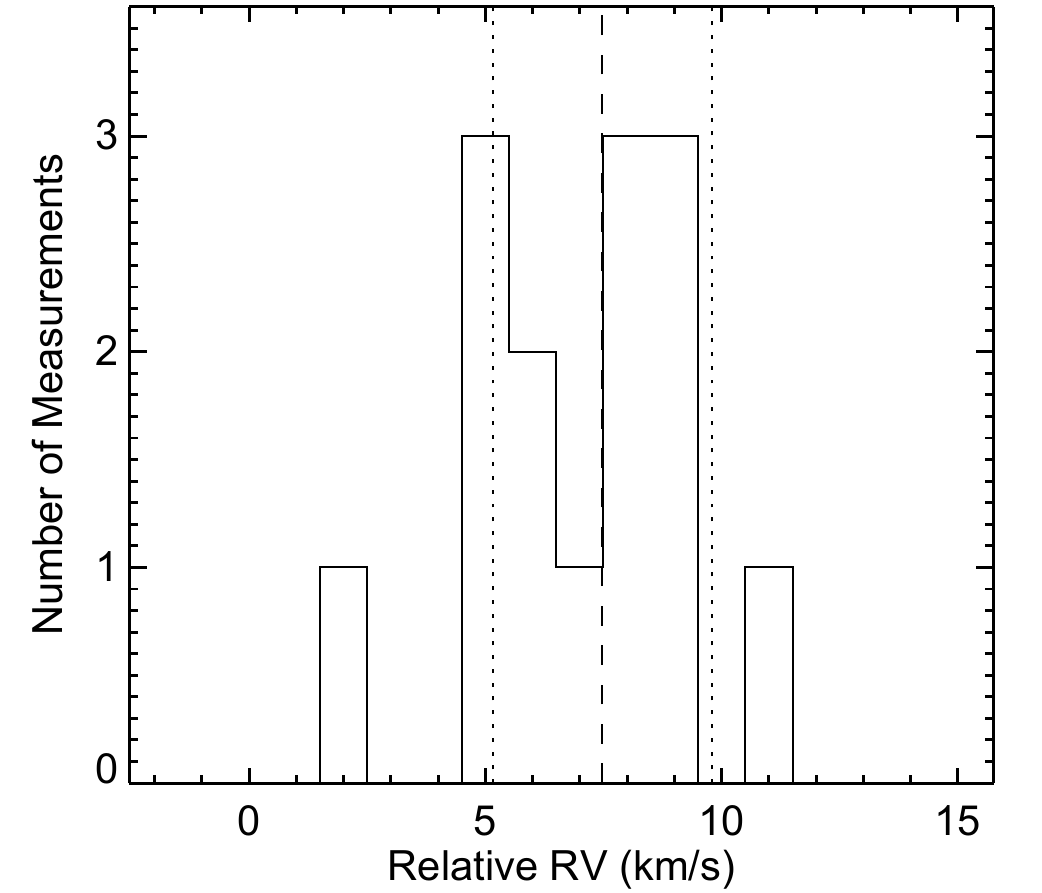}\label{fig:xcorlb}}
	\caption{Panel a:~$J$-band FIRE spectrum (black) of SDSS~J1110+0116 compared with the best-fitting BT-Settl model (red; \Teff\ = 1000\,K; $\log g$ = 4.5). The uncertainty spectrum is displayed in gray and the zero baseline as a dotted line. The \ion{K}{1} absorption doublet and features due to the CH$_4$ and H$_2$O features are identified.\\
	Panel b:~Histogram distribution of individual radial velocity measurements, compared with the adopted central value (dashed line) and the measurement uncertainties (dotted lines).
	\begin{center}(A color version of this figure is available in the online journal.)\end{center}}
	\label{fig:xcorl}
\end{figure}

We obtained a near-infrared (NIR) spectrum of SDSS~J1110+0116 with the Folded-port InfraRed Echellette (FIRE; \citealt{2013PASP..125..270S}) at the Magellan Telescopes on 2011 March 25 (UT), in conditions of light cirrus and 0$\farcs$7 seeing at $J$-band. We used the cross-dispersed echellette mode and 0$\farcs$6 slit to obtain 0.8--2.45\,$\mu$m spectroscopy at a resolving power $\lambda/\Delta\lambda$ $\approx$ 6000. Two exposures of 1204.7\,s were obtained at an airmass of 1.21--1.26. This was followed by observations of the A0~V star HD~93346 ($V$ = 7.42) at an airmass of 1.24 for telluric correction and flux calibration, and ThAr emission lamps for wavelength calibration. We obtained high- and low-illumination flat fields at the beginning of the night for pixel response calibration. All data were reduced using the Interactive Data Language (IDL) pipeline FIREHOSE, which is based on the MASE \citep{2009PASP..121.1409B} and SpeXTool \citep{2003PASP..115..389V,2004PASP..116..362C} packages, as described in \citet{2011AJ....142..169B}. The reduced spectrum, a portion of which is shown in Figure~\ref{fig:xcorla}, has a peak signal-to-noise of 30 per resolution element in the 1.2--1.3~$\mu$m region.

\section{RESULTS AND DISCUSSION}\label{sec:results}

\subsection{Radial Velocity Measurement}\label{sec:rvmes}

The radial velocity of SDSS~J1110+0116 was measured by cross-correlating its spectrum with zero-velocity BT-Settl models \citep{2012RSPTA.370.2765A} smoothed to the resolution of the FIRE data using a Gaussian profile. We first determined the best model parameters by comparing FIRE data to models with \Teff\ = \{800, 900, 1000, 1100, 1200, 1300\}\,K and $\log g$ = \{4.0, 4.5, 5.0\} (in units of \cmss) in the 1.20--1.33\,${\mu}$m region. The best-fitting model, \Teff\ = 1000\,K and $\log g$ = 4.5 is shown in Figure~\ref{fig:xcorla}. Note that these parameters are consistent with an age $<$\,250\,Myr (e.g., \citealt{2003A&A...402..701B}). 

We cross-correlated the model and observed spectrum over discrete spectral segments within the highly structured 1.26--1.31\,$\mu$m H$_2$O and CH$_4$ absorption region (Figure~\ref{fig:xcorl}a). As discussed in A. J. Burgasser et al. (in preparation), this region is particularly well-matched to spectral models and its structure allows fine sampling of wavelength shifts. We fit a total of 15 0.02\,$\mu$m regions in the 1.26--1.31\,$\mu$m range in 0.002\,$\mu$m (10 resolution elements) steps. This was done to account for small systematic effects that may be present in the wavelength calibration at the sub-pixel scale. Rejecting one outlier measurement, we find a mean and standard deviation in cross-correlation shifts of $7.5 \pm 2.3$\,\kms\ (Figure~\ref{fig:xcorl}b). Similar analysis of T dwarfs with radial velocity measurements (e.g., \citealt{2007ApJ...666.1205Z}; A. J. Burgasser et al., in preparation) suggests a 3\,\kms\ systematic uncertainty on this method, so we adopt a final radial velocity of $7.5 \pm 3.8$\,\kms. Given the low signal-to-noise spectrum, we did not attempt to measure $v\sin{i}$ for this source.

\subsection{AB~Doradus Membership}\label{sec:abdor}


Adding in the radial velocity measurement, we obtain a bayesian probability of 96.8\% that SDSS~J1110+0116 is a member of the AB~Doradus moving group. This is associated with a false positive field contamination probability of 0.4\%. We find a galactic position $XYZ$ = \{$-2.79 \pm 0.06$,$-10.78 \pm 0.25$,$15.62 \pm 0.36$\}\,pc and a space velocity $UVW$ = \{$-5.9 \pm 0.7$,$-30.1 \pm 2.2$,$-12.6 \pm 3.1$\}\,\kms\ for SDSS~J1110+0116, placing it $34 \pm 17$\,pc and $3.3 \pm 2.5$\,\kms\ away from the core of our AB~Doradus spatial and kinematic models, which is within the spread of members of this moving group. Based on this high membership probability, signs of low-gravity in its NIR spectrum, and the age estimate of 100--300\,Myr based on comparison to atmosphere models, we conclude that SDSS~J1110+0116 is a bona fide member of AB~Doradus.

\subsection{The Properties of SDSS~J1110+0116}\label{sec:properties}

\begin{deluxetable}{ll|ll}
\tablecolumns{4}
\tablecaption{Properties of SDSS~J1110+0116 \label{tab:properties}}
\tablehead{\colhead{Property} & \colhead{Value} & \colhead{Property} & \colhead{Value}}
\startdata
R.A. & 11:10:10.011 & $J$ (\emph{UKIDSS}) & $16.16 \pm 0.01$\\
Decl. & +01:16:13.09 & $H$ (\emph{UKIDSS}) & $16.20 \pm 0.02$\\
$\mu_\alpha\cos\delta$ (mas yr$^{-1}$)\tablenotemark{a} & $-217.1 \pm 0.7$ & $K$ (\emph{UKIDSS}) & $16.05 \pm 0.03$\\
$\mu_\delta$ (mas yr$^{-1}$)\tablenotemark{a} & $-280.9 \pm 0.6$ & $W1$ (\emph{AllWISE}) & $15.44 \pm 0.04$\\
RV (\kms) & $7.5 \pm 3.8$ & $W2$ (\emph{AllWISE}) & $13.92 \pm 0.04$\\
Distance (pc)\tablenotemark{a} & $19.19 \pm 0.44$ & $W3$ (\emph{AllWISE}) & $12.00 \pm 0.29$\\
$X$ (pc) & $-2.79 \pm 0.06$ & Spectral type & T5.5\,pec\\
$Y$ (pc) & $-10.78 \pm 0.25$ & Mass (\MJup) & 10--12\\
$Z$ (pc) & $15.62 \pm 0.36$ & Radius (\RJup) & $1.18 \pm 0.02$\\
$U$ (\kms) & $-5.9 \pm 0.7$ & \Teff\ (K) & $940 \pm 20$\\
$V$ (\kms) & $-30.1 \pm 2.2$ & $\log g$ & $4.28 \pm 0.04$\\
$W$ (\kms) & $-12.6 \pm 3.1$ & $\log{L_*/L_\odot}$ & $-4.99 \pm 0.02$\\[-5pt]
\enddata
\tablenotetext{a}{Measurement from \cite{2012ApJS..201...19D}}
\end{deluxetable}

We used the method described in Paper~II and Paper~V to estimate the mass of SDSS~J1110+0116; the method consists of comparing the absolute \emph{2MASS} $J$, $H$ and $K_S$ and WISE $W1$ and $W2$ magnitudes with the BT-Settl isochrones \citep{2003A&A...402..701B,2012RSPTA.370.2765A} at the age of AB~Doradus (110--130\,Myr; \citealt{2013ApJ...766....6B}) in a maximum likelihood analysis. We obtain a mass estimate of 10--11\,\MJup.

It is important to note that current evolution and atmosphere models likely suffer from important systematics, which are not well characterized and thus not included in our mass measurement error. For example, these models do not account for magnetic fields and assume a "hot start" formation, which could lead to an underestimation of the mass \citep{2014ApJ...792...37M,2014MNRAS.437.1378M}. Additionally, it has been demonstrated that current atmosphere models do not correctly reproduce the effects of dust, particularly in the case of young objects (e.g., \citealp{2011ApJ...735L..39B,2012ApJ...752...56F,2013ApJ...777L..20L}).

It has been proposed that the effects of enhanced dust clouds in young brown dwarfs also slow down the NIR cooling rates which would lead to an over-estimation of the mass \citep{2008ApJ...689.1327S}. This has been demonstrated by the recent dynamical mass measurements of dusty brown dwarfs (e.g., \citealp{2010ApJ...711.1087K,2015arXiv150306212D}).

We have furthermore used the method of J. Filippazzo et al. (submitted to ApJ) to obtain several of the fundamental parameters of SDSS~J1110+0116. We calculated the bolometric luminosity of SDSS~J1110+0116 using its parallax, NIR spectrum combined with NIR and mid-infrared photometry (see Table~\ref{tab:properties}) and obtained $\log{L_*/L_\odot} = -4.99 \pm 0.02$. We have then used solar metallicity, hybrid cloud (SMHC08) evolutionary model isochrones \citep{2008ApJ...689.1327S} to obtain a semi-empirical radius of $R = 1.18 \pm 0.02$\,\RJup\ using the above bolometric luminosity at the age of AB~Doradus (110--130\,Myr). This in turn allowed us to derive an effective temperature of \Teff\ $= 940 \pm 20$\,K from the Stefan-Boltzmann Law. Evolutionary tracks also provide estimates for the surface gravity ($\log g = 4.28 \pm 0.04$) and mass (10--12\,\MJup) of SDSS~J1110+0116, the latter of which is consistent with our previous mass estimate based on the BT-Settl models.

We used the same method to calculate the properties of all brown dwarf bona fide or high-confidence candidate members of AB~Doradus from the literature (\citealp{2012AJ....143...80S,2013AN....334...85L,2015ApJ...799..203G}; J. Gagn\'e et al., submitted to ApJ; J. K. Faherty et al., in preparation; see Table~\ref{tab:abdor}). In this work, the term \emph{high-confidence candidate member} refers to those that have at least an RV or a parallax measurement and that have unambiguous, high ($>$\,90\%) bayesian membership probabilities to AB~Doradus, as calculated by BANYAN~II.

We note that SDSS~J1110+0116 has an effective temperature which is marginally lower than those of typical T5--T6 dwarfs in the field (1000--1100\,K; \citealt{2009ApJ...702..154S}), as is the case for GU~Psc~b (see also \citealt{2014ApJ...787....5N}) and for the young ($\sim$\,125--400\,Myr) T2.5 dwarf HN~Peg~B \citep{2007ApJ...654..570L}. A similar but more pronounced effect has already been noted for young L-type brown dwarfs \citep{2006ApJ...651.1166M,2012ApJ...752...56F,2013AN....334...85L,2013ApJ...774...55B,2014ApJ...787....5N,2015ApJ...804...96G}. Table~\ref{tab:abdor} provides a sequence of the fundamental properties of young brown dwarfs at the age of AB~Doradus (110--130\,Myr) and how those relate to their NIR spectral types.

We note the interesting fact that 2MASS~J14252798--3650229 and 2MASS~J03552337+1133437 have remarkably similar fundamental physical properties albeit different NIR spectral shapes (e.g., J0355 has a much redder continuum, weaker FeH absorption at $\sim$\,1.0\,$\mu$m and a shallower CO band at $\sim$\,2.3\,$\mu$m). This is an additional indication for the diversity of the spectral properties of young brown dwarfs with similar fundamental physical properties (see \citealt{2013ApJ...772...79A}), an effect that could be explained by different cloud properties. Alternatively, this could also be a consequence of either object having an unresolved companion.


\begin{figure*}
	\centering
	\includegraphics[width=0.945\textwidth]{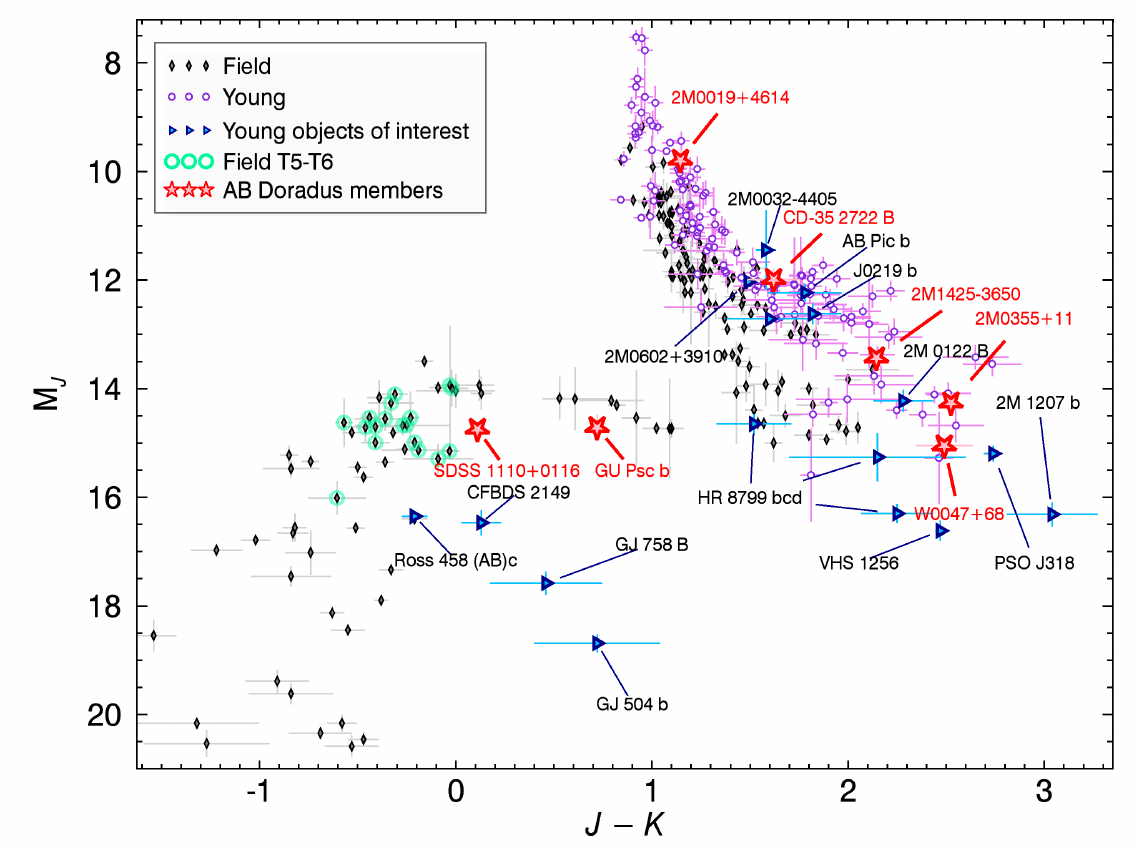}
	\caption{Color-magnitude diagram of field (black diamonds) and young (purple open circles) low-mass stars and brown dwarfs compared with the bona fide or high confidence brown dwarf members of AB~Doradus (red stars). Field dwarfs with spectral types in the T5--T6 range are circled in green for comparison with SDSS~J1110+0116. Young directly imaged planets, substellar companions and isolated brown dwarfs are displayed as blue right-pointing triangles for comparison. The NIR colors of SDSS~J1110+0116 are unusually red compared with field dwarfs of similar spectral types, despite its normal absolute $J$-band magnitude. $J$ and $K$ magnitudes are displayed in the Mauna Kea Observatory (MKO) system.
	\begin{center}(A color version of this figure is available in the online journal.)\end{center}}
	\label{fig:CMD}
\end{figure*}

In Figure~\ref{fig:CMD}, we compare the position of SDSS~J1110+0116 with that of field and young brown dwarfs and planetary-mass companions in a $M_J$ versus $J-K$ color-magnitude diagram. We display bona fide and high-confidence candidate members of AB~Doradus that are listed in Table~\ref{tab:abdor} using red star symbols to outline a preliminary sequence of brown dwarfs in AB~Doradus. These objects fall on the right of the field sequence, an effect that is also observed for earlier-type young brown dwarfs and planetary-mass companions (e.g., \citealp{2006ApJ...651.1166M,2008ApJ...689.1295K,2010ApJ...725.1405B,2011ApJ...735L..39B,2013AN....334...85L,2013AJ....145....2F}). We note that SDSS~J1110+0116 has absolute magnitudes similar to field T5--T6 dwarfs in the \emph{2MASS} $J$, $H$, $K_S$ and \emph{WISE} $W1$ and $W2$ bands \citep{2012ApJS..201...19D}. This may reflect a balance between a large radius and enhanced dust opacity in its high atmosphere. A compilation of the properties of SDSS~J1110+0116 are listed in Table~\ref{tab:properties}.

\subsection{The Search for a Co-Moving Companion}\label{sec:comoving}

We performed a search for a co-moving companion to SDSS~J1110+0116 using all 335 \emph{2MASS} entries within a conservatively large radius of 15$^\prime$, which corresponds to $\sim$\,17\,000\,AU at the distance of SDSS~J1110+0116. We cross-matched every \emph{2MASS} source with the \emph{AllWISE} catalog using the method described in Paper~V. The proper motions that we derived for this set of objects have a median precision of $\sim$\,20\,\masyr\ for both $\mu_\alpha\cos\delta$ and $\mu_\delta$. We find no object matching the proper motion of SDSS~J1110+0116 within 15$^\prime$ and $<$\,240\,\masyr. We can thus reject the possibility of a common proper motion companion that would be bright enough to be detected in the \emph{2MASS} and \emph{AllWISE} catalogs. The faintest of these 335 objects has $J = 17.3$ and $W1 = 17.1$, and the completeness limits of \emph{2MASS} and \emph{AllWISE} are $J = 15.8$ \citep{2006AJ....131.1163S} and $W1 = 17.1$\footnote{See \url{http://wise2.ipac.caltech.edu/docs/release/allwise/expsup/sec2\_4a.html}}, respectively.

\begin{deluxetable*}{lllll|ccccc}
\tablecolumns{10}
\tablecaption{Properties of brown dwarf high-confidence candidates or bona fide members of AB~Doradus \label{tab:abdor}}
\tablehead{
\colhead{Name} &
\colhead{Bona} &
\colhead{} &
\colhead{Spectral} &
\colhead{} &
\colhead{\LBol\tablenotemark{b}} & 
\colhead{Radius\tablenotemark{b}} & 
\colhead{\Teff\tablenotemark{b}} & 
\colhead{$\log g$\tablenotemark{b}} & 
\colhead{Mass\tablenotemark{c}}\\ 
\colhead{} &
\colhead{fide\,?} &
\colhead{Ref.} &
\colhead{type\tablenotemark{a}} &
\colhead{Ref.} &
\colhead{$\log{L_*/L_\odot}$} &
\colhead{(\RJup)} &
\colhead{(K)} &
\colhead{} &
\colhead{(\MJup)}}\\
\startdata
2MASS~J00192626+4614078 & N & 1 & M8\,$\beta$ & 2 & $-2.75 \pm 0.08$ & $1.65 \pm 0.11$ & $2890 \pm 170$ & $4.93 \pm 0.04$ & 80--111\\
CD--35~2722~B & Y & 3 & L3\,$\beta$ & 2,4 & $-3.45 \pm 0.08$ & $1.27 \pm 0.02$ & $2200 \pm 100$ & $4.84 \pm 0.06$ & 39--51\\
2MASS~J14252798--3650229 & Y & 2 & L4\,$\gamma$ & 2 & $-4.04 \pm 0.01$ & $1.23 \pm 0.01$ & $1590 \pm 10$ & $4.64 \pm 0.04$ & 25--29\\
2MASS~J03552337+1133437 & Y & 5 & L3--L6\,$\gamma$ & 2 & $-4.06 \pm 0.03$ & $1.20 \pm 0.04$ & $1590 \pm 40$ & $4.67 \pm 0.05$ & 21--30\\
2MASS~J00470038+6803543 & Y & 6 & L6--L8\,$\gamma$ & 2 & $-4.38 \pm 0.03$ & $1.28 \pm 0.07$ & $1280 \pm 40$ & $4.39 \pm 0.15$ & 13--22\\
GU~Psc~b & N & 7 & T3.5 & 7 & $-4.88 \pm 0.06$ & $1.19 \pm 0.03$ & $1000 \pm 40$ & $4.31 \pm 0.06$ & 10--13\\
SDSS~J111010.01+011613.1 & Y & 8 & T5.5 & 9 & $-4.99 \pm 0.02$ & $1.18 \pm 0.02$ & $940 \pm 20$ & $4.28 \pm 0.04$ & 10--12 \\[-5pt]
\enddata
\tablenotetext{a}{NIR spectral types}
\tablenotetext{b}{Estimates of fundamental parameters are based on the method described in Filippazzo et al. (submitted to ApJ).}
\tablenotetext{c}{Mass estimates were derived from a comparison of absolute NIR magnitudes with both the BT-Settl and the SMHC08 model isochrones at the age of AB~Doradus (110--130\,Myr), as described in the text.}
\tablecomments{References to this Table~: \\
(1)~\citealt{2012AJ....143...80S}; (2)~J. Gagn\'e et al., submitted to ApJ; (3)~\citealt{2011ApJ...729..139W}; (4)~\citealt{2013ApJ...772...79A}; (5)~\citealt{2013AN....334...85L}; (6)~\citealt{2015ApJ...799..203G}; (7)~\citealt{2014ApJ...787....5N}; (8)~This work; (9)~\citealt{2004AJ....127.3553K}.
}
\end{deluxetable*}

\section{CONCLUSION}\label{sec:conclusion}

Using existing previously reported astrometry and a new radial velocity measurement coupled with low-gravity features in its atmosphere, we have determined that SDSS~J1110+0116 is a T5.5 bona fide member of AB~Doradus, with an estimated mass of $\sim$\,10--12\,\MJup. This is one of the coldest member of any young moving group identified so far and its relatively high brightness will make it useful to better understand how age and surface gravity shape the atmospheres of low-mass brown dwarfs and planets, influence evolution, and guide future searches for planetary-mass members of young moving groups. This new object falls into a region of the mass/age parameter space that we have only recently begun to explore. It is similar to GU~Psc~b \citep{2014ApJ...787....5N} albeit for the fact that it has a slightly lower mass/temperature and is isolated in space. It is also comparable to PSO~J318.5338-22.8603 \citep{2013ApJ...777L..20L}, at an older age and with a slightly larger mass, or to the young T dwarfs ROSS~458~C \citep{2010ApJ...725.1405B} at a younger age and warmer temperature. All data presented in this work can be found at \url{www.astro.umontreal.ca/\textasciitilde gagne} and in the Montreal Spectral Library, located at \url{www.astro.umontreal.ca/\textasciitilde gagne/MSL.php}.

\acknowledgments

The authors would like to thank the anonymous referee, Rebecca Oppenheimer, France Allard and Michael~C. Cushing for useful comments and discussions. This work was supported in part through grants from the Fond de Recherche Qu\'eb\'ecois - Nature et Technologie and the Natural Science and Engineering Research Council of Canada. This research has benefited from the SpeX Prism Spectral Libraries, maintained by Adam Burgasser at \url{http://pono.ucsd.edu/\textasciitilde adam/browndwarfs/spexprism}, as well as the M, L, T and Y dwarf compendium housed at \url{http://DwarfArchives.org} and maintained by Chris Gelino, Davy Kirkpatrick, and Adam Burgasser, whose server was funded by a NASA Small Research Grant, administered by the American Astronomical Society. We thank Gregory Mace for having compiled and updated a list of known L, T and Y dwarfs as part of his PhD thesis.

This research made use of: the SIMBAD database and VizieR catalog access tool, operated at the Centre de Donn\'ees astronomiques de Strasbourg, France; data products from the Two Micron All Sky Survey (\emph{2MASS}), which is a joint project of the University of Massachusetts and the Infrared Processing and Analysis Center (IPAC)/California Institute of Technology (Caltech), funded by the National Aeronautics and Space Administration (NASA) and the National Science Foundation; the Extrasolar Planets Encyclopaedia (\url{exoplanet.eu}), which was developed and is maintained by the exoplanet TEAM; data products from the \emph{Wide-field Infrared Survey Explorer} (\emph{WISE}), which is a joint project of the University of California, Los Angeles, and the Jet Propulsion Laboratory (JPL)/Caltech, funded by NASA; the NASA/IPAC Infrared Science Archive (IRSA), which is operated by JPL, Caltech, under contract with NASA; the Infrared Telescope Facility (IRTF), which is operated by the University of Hawaii under Cooperative Agreement NNX-08AE38A with NASA, Science Mission Directorate, Planetary Astronomy Program; and the Database of Ultracool Parallaxes maintained by Trent Dupuy. This paper includes data gathered with the 6.5 meter Magellan Telescopes located at Las Campanas Observatory, Chile (CNTAC program CN2013A-135).

\indent \emph{Facilities:} Magellan:Baade (FIRE)
\bibliographystyle{apj}


\end{document}